\documentclass[12pt]{article}
\usepackage[margin=1in]{geometry}
\usepackage{graphicx}
\usepackage{caption}
\usepackage{soul, color}
\usepackage{subcaption}
\usepackage{mathtools}
\usepackage{verbatimbox}
\usepackage{url}
\usepackage[square,numbers]{natbib}

\usepackage{verbatim}
\usepackage{indentfirst}
\usepackage{physics}
\usepackage{amsmath}
\usepackage{authblk}
\usepackage{physics}
\usepackage{amsmath, amsfonts, amssymb}
\usepackage{xcolor}
\usepackage[]{hyperref}
\hypersetup{
	colorlinks=true,
	linkcolor={black},
	citecolor={cyan!80!black},
	filecolor={red!60!green},      
	urlcolor={magenta!60!black},
}

\usepackage[export]{adjustbox}
\usepackage{fix-cm}    
\makeatletter
\newcommand\HUGE{\@setfontsize\Huge{35}{40}}
\makeatother

\title{Path integrals, spontaneous localization, and the classical limit}

\author[1]{Bhavya Bhatt \thanks{coder1704@gmail.com}}
\author[2]{Manish Ram Chander\thanks{physicophillic@gmail.com}}
\author[3]{Raj Patil \thanks{patil.raj@students.iiserpune.ac.in}}
\author[4]{Ruchira Mishra \thanks{ruchiramishra98@gmail.com}}
\author[5]{Shlok Nahar \thanks{naharshlok@gmail.com}}
\author[6]{Tejinder P. Singh \thanks{tpsingh@tifr.res.in}}
\affil[1]{Indian Institute of Technology Mandi, Mandi 175005, India}
\affil[2]{Indian Institute of Technology Madras, Chennai 600036, India}
\affil[3]{Indian Institute of Science Education and Research Pune, Pune 411008, India}
\affil[4]{Indian Institute of Science Education and Research Mohali, Mohali 140306, India}
\affil[5]{Indian Institute of Technology Bombay, Mumbai 400076, India}
\affil[6]{Tata Institute of Fundamental Research, Homi Bhabha Road, Mumbai 400005, India}

\date{\today}

\begin{document}
\maketitle

\begin{abstract}

\noindent The measurement problem and the absence of macroscopic superposition are two foundational problems of quantum mechanics today. One possible solution is to consider the Ghirardi-Rimini-Weber (GRW) model of spontaneous localisation. Here we describe how spontaneous localisation modifies the path integral formulation of density matrix evolution in quantum mechanics. We provide two new pedagogical derivations of the GRW propagator. We then show how the von Neumann equation and the Liouville equation for the density matrix arise in the quantum and classical limit, respectively, from the GRW path integral.
\end{abstract}

\newpage

\tableofcontents
\newpage

\section{Introduction}
Non-Relativistic Quantum Mechanics is a general framework for all systems moving at speeds negligible in comparison to the speed of light. The theory is immensely successful for having predicted phenomena which have been {experimentally} verified extensively in the past hundred years. {However, there are some fundamental questions about the theory that still remain unanswered.} 

{One} of these is the measurement problem, which is essentially about how the collapse of a state occurs, and why is the outcome given by the Born rule \cite{wheeler2014quantum}. Yet another is the absence of macroscopic superpositions. QM predicts that any system can be in a superposition of ``states", but strangely, the effect is not easy to see at large length scales. Today, there is {an} ongoing effort {to} study macroscopic superposition{s} experimentally as well as theoretically \cite{emary2013leggett}.

Several proposals have been suggested to address these problems. However, the only experimentally verifiable modifications are the spontaneous collapse theories \cite{Pearle:76,GRW_CSL} where measurements are not considered special acts and are instead built into the evolution of the state. These theories are being experimentally tested by measuring the excess energy produced due to spontaneous localisation \cite{heateff, cmosc, blkheat}. Recently, an anomalous energy gain was detected using ultra-cold cantilevers \cite{cantilever} whose origin remains to be understood. We look at the simplest of these, the Ghirardi-Rimini-Weber (GRW) model {of quantum mechanics}, from the path integral perspective.

First derived by Pearle  and Soucek \cite{Pearle} in an alternative way, the GRW propagator is a generalization of the Feynman propagator and accounts for the pertinent phenomenological modifications. In this paper, we present two {pedagogical} {derivations of} this propagator, which {we believe} would be new additions to the literature. In particular, as we will see the correction to the standard propagator amounts to adding a damping term to it. This has possible repercussions for applications of path integral to quantum field theory. In addition, these methods can easily be extended to systems obeying the Lindblad equation which is ubiquitous in the study of open quantum systems as the GRW master equation is in Lindblad form. Thus, the applicability of this paper is broader than just the GRW model and it could improve our understanding of systems obeying the laws of standard quantum mechanics as well. We hope that our derivations would serve as an instructive source for the interested reader, beyond being a useful addition to the growing literature of collapse models.

\subsection{Introducing the model}

The idea of spontaneous localization, and collapse models in general, has been extensively studied in recent years, as a possible approach to solve the quantum measurement problem,  and explain the absence of macroscopic 
position superpositions. This was first proposed by Pearle in the
1970s \cite{Pearle:76} and subsequently by other authors in \cite{Ghirardi:86} and generalised to the case of identical
particles in the CSL model \cite{Ghirardi2:90}. The proposal is that every quantum object in nature undergoes spontaneous localisation to a region of size $r_c$, at random times given by a Poisson process with a mean collapse rate $\lambda$. Between every two collapses, the wave function obeys Schr\"{o}dinger evolution. The collapse rate can be shown to be proportional to the number $N$ of nucleons in the object, and we write $\lambda = N\lambda_{\text GRW}$, where $\lambda_{\text GRW}$  is the collapse rate for a nucleon. Thus, $\lambda_{\text GRW}$ and $r_C$ are two new constants of nature, whose values must be fixed by experiment.
Formally, the two postulates of the GRW model are stated as follows:

\smallskip

Postulate 1. Given the wave function $\psi ({\bf x_1}, {\bf x_2}, ..., {\bf x_N})$ of an $N$ particle quantum system in the Hilbert space {$\mathcal{L}^2(\vb{R}^{3N})$}, the $n$-th particle undergoes  spontaneous localization to a random position ${\bf x}$ as described by the following jump operator:
\begin{eqnarray}
{\psi_{t}({\bf x}_{1}, {\bf x}_{2}, \ldots {\bf x}_{N}) \quad
	\longrightarrow \quad} 
\frac{L_{n}({\bf x}) \psi_{t}({\bf x}_{1},
	{\bf x}_{2}, \ldots {\bf x}_{N})}{\|L_{n}({\bf x}) \psi_{t}({\bf
		x}_{1}, {\bf x}_{2}, \ldots {\bf x}_{N})\|}
\end{eqnarray}

The jump operator $L_{n}({\bf x})$ is a linear operator which is defined to be the normalised Gaussian:
\begin{equation}
L_{n}({\bf x}) =
\frac{1}{(\pi r_C^2)^{3/4}} e^{- ({\bf
		\hat q}_{n} - {\bf x})^2/2r_C^2}
\end{equation}
Here, ${\bf \hat q}_{n}$ is the position operator for the $n$-th particle of the system, and the random variable ${\bf x}$ is the spatial position to which the jump occurs. $r_C$, which is the width of the Gaussian, is a new constant of nature.

The probability density for the $n$-th particle to jump to the position
${\bf x}$ is assumed to be given by:
\begin{equation}
p_{n}({\bf x}) \quad \equiv \quad \|L_{n}({\bf x}) \psi_{t}({\bf
	x}_{1}, {\bf x}_{2}, \ldots {\bf x}_{N})\|^2
\end{equation}
Also, it is assumed  that the jumps are distributed in time as
a Poissonian process with frequency $\lambda_{\text{\tiny GRW}}$. This is the second
new constant of nature, in the model.\\

For an unentangled wave function we may write $ \psi_{t}({\bf x}_{1}, {\bf x}_{2}, \ldots {\bf x}_{N}) = \prod_n \phi_n(\vb{x}_n)$, where $\phi_n(\vb{x}_n)$ is the wave function for $n^{th}$ particle. Therefore, we have 
\begin{equation}
  p_n(\vb{x})=  \int d^3 x_n |L_n(\vb{x})\phi_n(\vb{x}_n)|^2  = \int  d^3 x_n [l(\vb{x, x}_n)]^2 |\phi_n(\vb{x}_n)|^2
\end{equation}
where $l(\vb{x, x}_n)$ is the position representative of the operator $L_n(\vb{x})$, a Gaussian localised at $\vb{x}$. Because it is an operator on $n^{th}$ particle's space, integrals over all other degrees of freedom are trivial. Further note that,
\begin{equation}
    \int d^3x [l(\vb{x},\vb{x'})]^2=1
\end{equation}
This ensures that $\int p_n(\vb{x}) d^3x = 1$. On the surface of the definition, this may not be obvious. The result follows similarly for the entangled states.

Postulate 2. In between any
two successive jumps, the wave function evolves according to the
Schr\"odinger equation.\\

{With these postulates, we can calculate the evolution of the density matrix that represents the state of the system as \cite{GRW_CSL}
\begin{equation}
    \frac{d}{dt}\rho(t)=-\frac{i}{\hbar}[H,\rho(t)]-\lambda(\rho(t)-\int d^3x L_n(x)\rho L_n(x))
\end{equation}
This can be rewritten in Lindblad form as
\begin{equation}
    \frac{d}{dt}\rho(t)=-\frac{i}{\hbar}[H,\rho(t)]+\int d^3x \lambda(L_n(x)\rho L_n(x)-\frac{1}{2}\{ [L_n(x)]^2,\rho(t)\})
\end{equation}
Thus, the GRW equation is an example of a Lindblad equation and the following methods to derive the path integral can also be used to derive the path integral for any open quantum system satisfying the Lindblad master equation.}

{For the above model,} the process of spontaneous localisation serves to provide an exponential damping of the exponential oscillations in the path integral amplitude. Inevitably, the damping is important for macroscopic systems, but insignificant for microscopic ones.

\section{The GRW path integral and its derivation}
The path integral formulation of quantum mechanics is a description of quantum theory that generalizes the action principle of classical mechanics. It replaces the classical notion of a single, unique classical trajectory for a system with a sum, or functional integral, over an infinity of quantum-mechanically possible trajectories to compute a quantum amplitude. As mentioned in the introduction, the GRW path integral has been previously derived by Pearle and Soucek \cite{Pearle}; here we give two alternative derivations of their result, and then discuss the classical and quantum limits of the GRW path integral. [For further applications of path integrals to collapse models, see also \cite{B1,B2,B3}].

\subsection{Method-1}
\subsubsection{Introduction}
Standard techniques \cite{RShankar} can be used to derive the propagator  starting from the Schr\"odinger equation. However, these techniques cannot directly be used for mixed states represented by density matrices. Hence, we first purify the state-vector \cite{dentoket} so that it obeys  Schr\"odinger-like evolution with an effective Hamiltonian. The methods followed in \cite{RShankar} can then be directly applied to this pure state ket. {Such a method of purification of a density matrix to simplify its treatment can also be used when deriving the path integral of a more general class of open quantum system as described in the Introduction as this method does not depend on any property of the GRW equation other than the fact that it preserves trace.}

\subsubsection{Getting the Hamiltonian Form}The GRW master equation {for a single particle} \cite{Ghirardi2:90, GRW_CSL} is
\begin{equation} \label{GRW}
\frac{d\rho}{dt}= -\frac{i}{\hbar}(H\rho-\rho H)-\lambda\left(\rho-\int d^3r L_r \rho L_r\right)
\end{equation}
where $H$ is the Hamiltonian for Schr\"odinger evolution of the system and \begin{equation}\label{jump}
L_r=\frac{1}{\mathcal{N}} \exp\left({-\frac{(\hat{q}-r)^2}{2r_C^2}}\right)
\end{equation} is the collapse operator for the particle to localize around $r$. $\lambda$ is the collapse rate, and $r_C$ is the length scale to which localization takes place, as defined in the introduction. This master equation was first derived for the CSL model \cite{Ghirardi2:90} where the authors noted that for the one particle case this equation is the same as for the GRW model, although this is not true in general.

In order to convert Eq. (\ref{GRW}) into an equation of the form
\begin{equation} \label{Sc}
\frac{d\ket{\psi}}{dt}= -\frac{i}{\hbar}\tilde{H}\ket{\psi}
\end{equation}
we define $\ket{\psi}$ as
\begin{equation} \label{psi}
\ket{\psi}=\sum_{m,n}\rho_{mn} \ket{m}\otimes \ket{n}
\end{equation}
where $\rho_{mn}=\bra{m}\rho \ket{n}$ are elements of the density matrix $\rho$ from Eq. (\ref{GRW}). {Here the set of all $\ket{m}, \ket{n}$ form an orthonormal basis in the single particle Hilbert space}. We note that there is an isomorphism between $\ket{\psi}$ as defined here, and $\rho$. Thus, knowing the evolution of $\ket{\psi}$ would give us all the information about how $\rho$ whould evolve. Using Einstein's summation convention, we rewrite Eq. (\ref{GRW}) as,
\begin{equation} \label{index1}
\frac{d\rho_{mn}}{dt}= -\frac{i}{\hbar}(H_{ma}\rho_{an}-\rho_{ma} H_{an})-\lambda\left(\rho_{mn}-\int d^3r {L_r}_{ma} \rho_{ab} {L_r}_{bn}\right)
\end{equation}
From Eq. (\ref{Sc}) and Eq. (\ref{psi}), it follows that the equation 
\begin{equation} \label{index2}
\frac{d\rho_{mn}}{dt}= -\frac{i}{\hbar}\tilde{H}_{mabn}\; \rho_{ab}
\end{equation} must also hold.
Comparing Eq. (\ref{index1}) and Eq. (\ref{index2}) we get 
\begin{equation}
\tilde{H}=(H\otimes \mathbb{I}-\mathbb{I}\otimes H^\mathsf{T}) - (i\hbar \lambda \mathbb{I} \otimes \mathbb{I})+i\hbar \lambda \int d^3r L_r \otimes L_r^\mathsf{T}
\end{equation}
So, $\ket{\psi (t)}$ evolves as \begin{equation}\label{rhoev}
\ket{\psi (t)}=\exp({-i\tilde{H}t/\hbar})\ket{\psi (0)}
\end{equation} 
This gives us the evolution of $\rho(t)$ via Eq. (\ref{psi}), and the above equation can be used to derive the propagator and the path integral.

\subsubsection{Derivation of the Path Integral}
The total time $t=T$ can be divided into $N$ intervals such that $\epsilon=T/N$ and the finite time propagator in Eq. (\ref{rhoev}) can be written as
\begin{equation}
U=\left[\exp\left(\frac{-i\epsilon}{\hbar}\left(H\otimes \mathbb{I}-\mathbb{I}\otimes H^\mathsf{T}\right) - \lambda \epsilon\left(\mathbb{I} \otimes \mathbb{I} - \int d^3r L_r \otimes L_r^\mathsf{T}\right)\right)\right]^N
\end{equation}
As $N \rightarrow \infty$ and $\epsilon \rightarrow 0$, we can make the approximation 
\begin{equation}
U \approx \left[\exp\left(\frac{-i\epsilon}{\hbar}(H\otimes \mathbb{I}-\mathbb{I}\otimes H^\mathsf{T})\right)\times \exp\left(- \lambda \epsilon(\mathbb{I} \otimes \mathbb{I} - \int d^3r L_r \otimes L_r^\mathsf{T})\right)\right]^N
\end{equation}
Introducing resolution of the identity \begin{equation}\label{key}
\int_{-\infty}^{\infty}dx_{k\epsilon}dy_{k\epsilon} \ket{x_{k\epsilon}}\ket{y_{k\epsilon}}\bra{x_{k\epsilon}}\bra{y_{k\epsilon}}
\end{equation} between every time step we get $N$ terms, each of the form \begin{equation}\label{key}
\bra{x_{k\epsilon},{y_{k\epsilon}}}\exp\left[\frac{-i\epsilon}{\hbar}(H\otimes \mathbb{I}-\mathbb{I}\otimes H^\mathsf{T})\right]\times \exp\left[- \lambda \epsilon\left(\mathbb{I} \otimes \mathbb{I} - \int d^3r L_r \otimes L_r^\mathsf{T}\right)\right]\ket{x_{(k-1)\epsilon},{y_{(k-1)\epsilon}}}
\end{equation} Evaluating one of these terms
\begin{align}\label{key}
&\bra{x_{k\epsilon},{y_{k\epsilon}}}\exp\left[\frac{-i\epsilon}{\hbar}\left(H\otimes \mathbb{I}-\mathbb{I}\otimes H^\mathsf{T}\right)\right] \times \exp\left[- \lambda \epsilon\left(\mathbb{I} \otimes \mathbb{I} - \int d^3r L_r \otimes L_r^\mathsf{T}\right)\right]\ket{x_{(k-1)\epsilon},{y_{(k-1)\epsilon}}} \nonumber
\\
&= \bra{x_{k\epsilon},{y_{k\epsilon}}}\exp\left[\frac{-i\epsilon}{\hbar}\left(H\otimes \mathbb{I}-\mathbb{I}\otimes H^\mathsf{T}\right)\right]\ket{x_{(k-1)\epsilon},{y_{(k-1)\epsilon}}}\exp\left[-\lambda \epsilon\left(1 - \exp\frac{-(x_{(k-1)\epsilon}-y_{(k-1)\epsilon})^2}{4r_C^2}\right)\right]
\end{align}

The first exponent is simply the Feynman propogator\footnote{The propagator is a function that specifies the probability amplitude for a particle to travel from one place to another in a given time, or to travel with a certain energy and momentum.} for Schr\"{o}dinger evolution.\footnote{Refer R. Shankar \cite{RShankar}; Eq. (21.1.15)} {We now assume that the system is non relativistic and hence, Hamiltonian is quadratic in the momentum. and the potential is position dependent.} Thus, taking all $N$ terms we get
\begin{align}
U(x_{N\epsilon},y_{N\epsilon},x_0,y_0)&=\frac{m}{2\pi \epsilon \hbar}\times\int_{-\infty}^{\infty}\prod_{n=1}^{N-1}\frac{m}{2\pi \epsilon \hbar}dx_{n\epsilon}dy_{n\epsilon}\times\exp\Big[\sum_{n=1}^{N}\left(\frac{im}{2\hbar \epsilon}((x_{n\epsilon}-x_{(n-1)\epsilon})^2-(y_{n\epsilon}-y_{(n-1)\epsilon})^2)\right)\nonumber\\&~~~~~~~-\frac{i\epsilon}{\hbar}\left(V(x_{(n-1)\epsilon})-V(y_{(n-1)\epsilon})\right)\Big] \times\exp\left[\sum_{k=1}^{N} -\lambda \epsilon\left(1 - \exp\frac{-(x_{(k-1)\epsilon}-y_{(k-1)\epsilon})^2}{4r_C^2}\right)\right]
\\&=\frac{m}{2\pi \epsilon \hbar}\int_{-\infty}^{\infty}\prod_{n=1}^{N-1}\frac{m}{2\pi \epsilon \hbar}dx_{n\epsilon}dy_{n\epsilon}\times\exp\left[\sum_{n=1}^{N}\frac{i}{\hbar}(S[x_{n\epsilon},x_{(n-1)\epsilon}]-S[y_{n\epsilon},y_{(n-1)\epsilon}])\right]\nonumber\times
\\&~~~~~~\exp\left[\sum_{k=1}^{N} -\lambda \epsilon\left(1 - \exp\frac{-(x_{(k-1)\epsilon}-y_{(k-1)\epsilon})^2}{4r_C^2}\right)\right]
\end{align}
In the continuum limit with $N \rightarrow \infty$ while still keeping $N \epsilon=T$, the evolution of the density matrix element thus becomes,
\begin{align}\label{eq_path_int_shlok}
\rho(x_T,y_T,T)=&\int_{\rm all\  paths} \mathcal[{D}x_t]\mathcal[{D}y_t]\times \exp\left(\frac{i}{\hbar}(S[x_t,T,t=0]-S[y_t,T,t=0])\right)\times\nonumber\\&~~~~~~~~~~~~~~~~\exp\left[- \lambda\int_0^T dt\left(1 - \exp\frac{-
	(x_t-y_t)^2}{4r_C^2}\right)\right]\; \rho(x_0,y_0)dx_0dy_0
\end{align}
where \begin{equation}\label{eq_integral_def}
\int\mathcal[{D}x_t]=\lim\limits_{N \rightarrow \infty}{\left(\frac{m}{2\pi\hbar\epsilon}\right)}^{1/2}\int\prod_{n=1}^{N-1}{\left(\frac{m}{2\pi\hbar\epsilon}\right)}^{1/2}dx_n
\end{equation}
This is the same result as derived in \cite{Pearle}.
Here, $x_t$ and $y_t$ can be understood as individual paths that might be traversed. Thus, $\int_{\rm all \ paths} [Dx_t] [Dy_t]$ can be understood as an integral over all such paths.
The exponential in the second line of the above equation serves as the GRW induced regulator of the Feynman path integral, and improves the understanding of the classical limit, as we will see in the next section.

\subsection{Method-2}
\subsubsection{Introduction}
In this case, we use a more physically motivated approach. We use the fact that after every time interval $\epsilon$ the wave function has a probability $ \lambda\epsilon $ to collapse. Thus, by taking discrete time steps and using the above fact, we can derive the propagator. 
\subsubsection{Derivation of the Path Integral}
Consider $\rho(x_0,y_0,t=0)$ to be a density matrix at initial time $t=0$.  We intend to find $\rho(x_T,y_T,T)$ at final time $t=T$. We divide the total time into smaller intervals such that $\epsilon=\frac{T}{N}$. So, we have
\begin{align}\label{xx}
\rho(x_\epsilon,y_\epsilon,\epsilon)&=A\int_{-\infty}^{\infty}\int_{-\infty}^{\infty}\exp\left[\frac{i}{\hbar}\left(\frac{m}{2}(\frac{x_{\epsilon}-x_0}{\epsilon})^2-V(\frac{x_{\epsilon}+x_0}{2})\right)\epsilon\right]\nonumber\\&~~~~~~~~~~~~~~~~~~~\qquad \exp\left[-\frac{i}{\hbar}\left(\frac{m}{2}(\frac{y_{\epsilon}-y_0}{\epsilon})^2-V(\frac{y_{\epsilon}+y_0}{2})\right)\epsilon\right]\rho(x_0,y_0,t=0)d{x_0}d{y_0}\\&=A\int_{-\infty}^{\infty}\int_{-\infty}^{\infty}\mathcal{P}_1\rho(x_0,y_0,t=0)d{x_0}d{y_0}
\end{align} where
\begin{align}
\mathcal{P}_{i} &= K_{\rho}(x_{i}, y_{i}, \epsilon_{i}; x_{i-1}, y_{i-1}, \epsilon_{i-1}) \\&= \exp\left[\frac{i}{\hbar}\left(\frac{m}{2}(\frac{x_{i}-x_{i-1}}{\epsilon})^2-V(\frac{x_{i}+x_{i-1}}{2})\right)\epsilon\right]\nonumber\\&~~~~~~~~~~~~~~~~~~~\qquad \exp\left[-\frac{i}{\hbar}\left(\frac{m}{2}(\frac{y_{i}-y_{i-1}}{\epsilon})^2-V(\frac{y_{i}+y_{i-1}}{2})\right)\epsilon\right]
\end{align} 
is density matrix propagator for infinitesimal time step from $\epsilon_{i-1}$ to $\epsilon_{i}=\epsilon_{i-1} + \epsilon$. The above expression for $\rho(x_{\epsilon}, y_{\epsilon}, \epsilon)$ represents standard Schr\"odinger evolution\footnote{Refer R. Shankar \cite{RShankar} Eq. (8.5.4)} where $A$ is the appropriate normalization constant to recover Von-Neumann Equation. Now from Eq. (\ref{GRW}) we know that at a given instant say, $t=\epsilon$ the probability of collapse is $\lambda\epsilon$ while that of it evolving according to Schr\"odinger's equation is $1-\lambda\epsilon$.
Thus, the new density matrix after $ \epsilon  $ time becomes,
\begin{equation}
\rho_{new}(x_\epsilon,y_\epsilon,\epsilon)=(1-\lambda \epsilon)\rho_1+\lambda \epsilon \int_{-\infty}^{\infty}L_r(x_\epsilon)\rho_1 L_r(y_\epsilon)d{r}
\end{equation}where $\rho_1=\rho(x_\epsilon,y_\epsilon,\epsilon)$ and $L_r(x_\epsilon)=\bra{x_\epsilon}L_r\ket{x_\epsilon}$ are as defined in Eq. (\ref{jump}).
Here, since $\rho_1$ does not depend on $r$ (it is a function of $x_\epsilon,y_\epsilon$, $x_0$ and $y_0$), we can evaluate the above integral by taking $\rho_1$ outside the integration. We get,

\begin{align}
\int_{-\infty}^{\infty}L_r(x_\epsilon)\rho_1 L_r(y_\epsilon)d{r}& =\left(\int_{-\infty}^{\infty}L_r(x_\epsilon)L_r(y_\epsilon)d{r}\right)\rho_1 \\
& =\left[\int_{-\infty}^{\infty}\exp\left(-\frac{(x_\epsilon-r)^2}{2r_C^2}\right)\exp\left(-\frac{(y_\epsilon-r)^2}{2r_C^2}\right)d{r}\right]\rho_1 \\
& =\exp\left[-\frac{(x_\epsilon-y_\epsilon)^2}{4r_C^2}\right]\rho_1
\end{align}
{Now}, we can write \footnote{{Note that now the function in eq.(28) is commutative as it does not contain any differential operations}}
\begin{equation}
\rho_{new}(x_\epsilon,y_\epsilon,\epsilon)=(1-\lambda \epsilon)\rho_1+\lambda \epsilon\left[\exp\left(-\frac{(x_\epsilon-y_\epsilon)^2}{4r_C^2}\right)\right]\rho_1
\end{equation}
For simplicity we write
\[G_i=\exp\left[-\frac{(x_{i\epsilon}-y_{i\epsilon})^2}{4r_C^2}\right]\] and thus
\begin{equation}
\rho_{new}(x_\epsilon,y_\epsilon,\epsilon)=\left[(1-\lambda \epsilon)+\lambda \epsilon G_1\right]\rho_1
\end{equation}We propagate again according to Schr\"odinger's equation from time $t=\epsilon$ to time $t=2\epsilon$,
\begin{equation}\rho(x_{2\epsilon},y_{2\epsilon},2\epsilon)=A\int_{-\infty}^{\infty}\mathcal{P}_2\left[(1-\lambda \epsilon) + \lambda \epsilon G_1\right]\rho_1d{x_\epsilon}d{y_\epsilon}
\end{equation} 
Substituting $\rho_1$ according to the Eq. (\ref{xx}) and writing new $\rho_{new}(x_{2\epsilon},y_{2\epsilon},2\epsilon)$, we get
\begin{align}
\rho_{new}(x_{2\epsilon},y_{2\epsilon},2\epsilon)=&A^2((1-\lambda \epsilon)+\lambda \epsilon G_2)\int_{-\infty}^{\infty}\int_{-\infty}^{\infty}\mathcal{P}_2\left((1-\lambda \epsilon) + \lambda \epsilon G_1\right)\nonumber\\&~~~~~~~~~~~\int_{-\infty}^{\infty}\int_{-\infty}^{\infty}\mathcal{P}_1\rho(x_0,y_0,t=0)d{x_0}d{y_0}d{x_\epsilon}d{y_\epsilon}
\end{align} 
Further, {we can rearrange the terms as all the functions $\mathcal{P}_1, \mathcal{P}_2, G_1$ and $G_2$ are commutative operations so this gives us, }
\begin{align}
\rho_{new}(x_{2\epsilon},y_{2\epsilon},2\epsilon)=&A^2\int_{-\infty}^{\infty}\int_{-\infty}^{\infty}\int_{-\infty}^{\infty}\int_{-\infty}^{\infty}\mathcal{P}_2 \mathcal{P}_1\left((1-\lambda \epsilon)+\lambda \epsilon G_2\right)\left((1-\lambda \epsilon) + \lambda \epsilon G_1\right)\nonumber\\&~~~~~~~~~~~~~~~~~~~~~~~~~~~~~~~~~~~~~~~~\rho(x_0,y_0,t=0)d{x_0}d{y_0}d{x_\epsilon}d{y_\epsilon}
\end{align} 
We repeat the above procedure $N-1$ times. Taking continuum limit ${N\to\infty}$ gives us the final density matrix as
\begin{align}\label{xy}
\rho(x_T,y_T,T)=&\lim_{N\to\infty} A^{N-1}\idotsint\prod_{i=0}^{N-1}\mathcal{P}_i\prod_{i=0}^{N-1}\left((1-\lambda \epsilon)+\lambda \epsilon G_i\right)\nonumber\\&~~~~~~~~~~~~~~~~~~~~~~~~~~~~~~~~~\rho(x_0,y_0,t=0)d{x_0}d{y_0}\cdots d{x_{(N-1)\epsilon}}d{y_{(N-1)\epsilon}}
\end{align}
We know that
\begin{align}
\lim_{N\to\infty}\prod_{i=1}^{N}\mathcal{P}_i& =\lim_{N\to\infty}\prod_{i=1}^{N}\exp\left\{\frac{i}{\hbar}\left[\frac{m}{2}\left(\frac{x_{i\epsilon}-x_{(i-1)\epsilon}}{\epsilon}\right)^2-V\left(\frac{x_{i\epsilon}+x_{(i-1)\epsilon}}{2}\right)\right]\epsilon\right\}\nonumber\\&~~~~~~~~~~~~~~~~~~~~~~~~~~~~~~~~~~\exp\left\{-\frac{i}{\hbar}\left[\frac{m}{2}\left(\frac{y_{i\epsilon}-y_{(i-1)\epsilon}}{\epsilon}\right)^2-V\left(\frac{y_{i\epsilon}+y_{(i-1)\epsilon}}{2}\right)\right]\epsilon\right\} \\
& =\exp\left\{\lim_{N\to\infty}\sum_{i=1}^{N}\frac{i}{\hbar}\left[\frac{m}{2}\left(\frac{x_{i\epsilon}-x_{(i-1)\epsilon}}{\epsilon}\right)^2-V\left(\frac{x_{i\epsilon}+x_{(i-1)\epsilon}}{2}\right)\right]\epsilon\right\}\nonumber\\&~~~~~~~~~~~~~~~~~~~~~~~~~~~~~~~~~~\exp\left\{\lim_{N\to\infty}\sum_{i=1}^{N}-\frac{i}{\hbar}\left[\frac{m}{2}\left(\frac{y_{i\epsilon}-y_{(i-1)\epsilon}}{\epsilon}\right)^2-V\left(\frac{y_{i\epsilon}+y_{(i-1)\epsilon}}{2}\right)\right]\epsilon\right\} \\
& =\exp\left\{\frac{i}{\hbar}\int_{0}^{T}L\left(x_t\right)d{t}\right\}\exp\left\{-\frac{i}{\hbar}\int_{0}^{T}L\left(y_t\right)d{t}\right\} \\
& =\exp\left\{{\frac{i}{\hbar}(S(x_t,T,t=0)-S(y_t,T,t=0))}\right\}
\end{align}
{where, $ L\left(x_t,T,t=0\right) $ is the Lagrangian and $ S(x_t,T,t=0) $ the action thus obtained.
Expanding the second product term gives us,
\begin{align}
\lim_{N\to\infty}\prod_{i=1}^{N}\left((1-\lambda \epsilon)+\lambda \epsilon G_i\right)& =\lim_{N\to\infty}(1-\lambda \epsilon)^{N}\sum_{k=0}^{\infty}\frac{(\sum_{i=1}^{N}\lambda G_{i}\epsilon)^{k}}{k!(1 - \lambda\epsilon)^{k}} \\
& =\exp({-\lambda T})\exp\left(\lim_{N\to\infty}\sum_{i=1}^{N}\lambda G_i\epsilon\right)\\
& =\exp({-\lambda T})\exp\left(\lambda\int_{0}^{T}G(t)d{t}\right) \\
& =\exp({-\lambda T})\exp\left[\lambda\int_{0}^{T}\exp({-\frac{(x_t-y_t)^2}{4r_C^2}})d{t}\right]                                                                             
\end{align}
Substituting these two terms back in Eq. (\ref{xy}) we get an integral form solution of Eq. (\ref{GRW}) given by
\begin{align}\label{eq_path_int_bhavya}
\rho(x_T,y_T,T)=&\int_{}~~ \exp\left({\frac{i}{\hbar}(S[x_t,T,t=0]-S[y_t,T,t=0])}\right)\nonumber\\
&~~\exp\left(-\lambda\int_{0}^{T}(1-\exp\left\{{-\frac{(x_t-y_t)^2}{4r_C^2}}\right\})d{t}\right) [Dx_t][Dy_t]\rho(x_0,y_0,t=0)d{x_0}d{y_0}
\end{align}
where, the integral in the above equation is defined in the equation (\ref{eq_integral_def}). The above derived propagator is the same as what we got using the previous method given in equation (\ref{eq_path_int_shlok}). }

\section{Classical and Quantum Limits of GRW path integral}
\subsection{Quantum Limit}
{From equations (\ref{eq_path_int_shlok}) or (\ref{eq_path_int_bhavya}), the path integral for the GRW model is written as
\begin{align}
\label{eq_GRW_path_int_prop_1}
\rho(x_T,y_T,T)=&\int_{\rm all\ paths}\exp\left[\frac{i}{\hbar }\left( ~S(x_t,T,t=0) - S(y_t,T,t=0)~ \right)\right]\exp\left[- \lambda \int_{0}^{T} \left(1 - e^\frac{-(x_t-y_t)^2}{4r_C^2}\right)dt \right]\nonumber\\&~~~~~~~~~~~~~~~~~~~~~~~~~~~~~~~~~~~~~~~~~\rho(x_0,y_0,t=0) [Dx_t][Dy_t] dx_0 dy_0
\end{align}}
If we consider the limit $\lambda T \rightarrow 0$, i.e. we look at the system at timescales ($t = T$) much smaller than the time period of collapse ($ \tau = 1/\lambda $), then the non-oscillating part of the above given propagator could be approximated as,
\begin{equation}
\exp\left[- \lambda\int_0^T (1 - e^\frac{-(x_t-y_t)^2}{4r_C^2})dt \right] \approx 1
\end{equation}
This makes the propagator of GRW  look exactly like that for normal quantum mechanics,
\begin{align}\label{eq_GRW_infin_time_daig_terms_1}
\rho(x_T,y_T,T)=&\int_{\rm all\ paths}  \exp\left[\frac{i}{\hbar }\left( ~S(x_t,T,t=0) - S(y_t,T,t=0)~ \right)\right]\rho(x_0,y_0,t=0) [Dx_t][Dy_t] dx_0 dy_0
\end{align}
From here the standard quantum mechanical result follows easily - we recall the calculation here, for sake of completeness. We can write the above equation for infinitesimal time interval $\epsilon$ as,
\begin{align}
\rho(x_\epsilon,y_\epsilon,\epsilon)&=A\int_{}~~\exp\left[\frac{i}{\hbar}\int_{0}^{\epsilon}\left(\frac{m\dot{x}^2}{2}-V(x)\right)dt\right]\nonumber\\
&~~~~~~~~~~~~~\exp\left[-\frac{i}{\hbar}\int_{0}^{\epsilon}\left(\frac{m\dot{y}^2}{2}-V(y)\right)dt\right]\rho(x_0,y_0,t=0)d{x_0}d{y_0}
\end{align}
where $A$ is as defined in the previous section. Using the following finite difference substitution
\[
\dot{x}\rightarrow \frac{x_{\epsilon}-x_0}{\epsilon}
\]
\[
x\rightarrow \frac{x_0+x_\epsilon}{2}
\]and using the standard substitution of $\eta_x = x_0 - x_\epsilon$ and $\eta_y = y_0 - y_\epsilon$ and rearranging the terms we have
\begin{equation}\label{eq_prop_1}
\rho(x_\epsilon,y_\epsilon,\epsilon)=A\int_{}\int_{}e^{\frac{i}{\hbar}\frac{m\eta_x^2}{2\epsilon}}e^{\frac{-i}{\hbar}\frac{m\eta_y^2}{2\epsilon}}\exp\left[\frac{i}{\hbar}\left(-V(x)+V(y)\right)\epsilon\right]\rho(x_\epsilon+\eta_x,y_\epsilon+\eta_y,t=0)d{\eta_x}d{\eta_y}
\end{equation}
The exponentials oscillate very rapidly as $\epsilon$ could be made arbitrarily small. When such a rapidly oscillating function
multiplies a smooth function, the integral vanishes for the most part due to the random phase of the exponential. Just as in the case of the path integration, the only substantial contribution comes from the region where the phase is stationary. The region of constructive interference is,
\begin{equation}
\frac{m \eta^2}{2 \hbar \epsilon} \leq \pi
\end{equation}
Now, Taylor expanding the terms in equation (\ref{eq_prop_1}) upto first order in $\epsilon$ i.e. upto order $\eta^2$ we get
\begin{equation}
\begin{split}
\rho(x_\epsilon,y_\epsilon,\epsilon)& = A\int_{}\int_{}e^{\frac{i}{\hbar}\frac{m\eta_x^2}{2\epsilon}}e^{\frac{-i}{\hbar}\frac{m\eta_y^2}{2\epsilon}}\left(1-\frac{i}{\hbar}V(x)\epsilon+\frac{i}{\hbar}V(y)\epsilon\right)\rho(x_\epsilon+\eta_x,y_\epsilon+\eta_y,t=0)d{\eta_x}d{\eta_y} \\
& = A \int_{}\int_{}e^{\frac{i}{\hbar}\frac{m\eta_x^2}{2\epsilon}}e^{\frac{-i}{\hbar}\frac{m\eta_y^2}{2\epsilon}}\left(1-\frac{i}{\hbar}V(x)\epsilon+\frac{i}{\hbar}V(y)\epsilon\right)\big(\rho(x_\epsilon,y_\epsilon,t=0)\\
&~~~~~~~~~~~~~~~~~~~~~~~~~~~~~~~~~~~~~~~~~~~~~+\frac{\partial\rho}{\partial y}\Big|_{(x_\epsilon,y_\epsilon,t=0)}\eta_y+\frac{\partial\rho}{\partial x}\Big|_{(x_\epsilon,y_\epsilon,t=0)}\eta_x \\&~~~~~~~~~~~~~~~~~~~~~~~~~~~~~~~~~~~~~~~~~~~~~+\frac{\partial^2\rho}{2\partial y^2}\Big|_{(x_\epsilon,y_\epsilon,t=0)}\eta_y^2+\frac{\partial^2\rho}{2\partial x^2}\Big|_{(x_\epsilon,y_\epsilon,t=0)}\eta_x^2\big)d{\eta_x}d{\eta_y}\\
& = A \int_{}\int_{}e^{\frac{i}{\hbar}\frac{m\eta_x^2}{2\epsilon}}e^{\frac{-i}{\hbar}\frac{m\eta_y^2}{2\epsilon}}\big(\rho(x_\epsilon,y_\epsilon,t=0)-\frac{i}{\hbar}V(x)\rho(x_\epsilon,y_\epsilon,t=0)\epsilon\\
&~~~~~~~~~~~~~~~~~~~~~+\frac{i}{\hbar}V(y)\rho(x_\epsilon,y_\epsilon,t=0)\epsilon+\frac{\partial^2\rho}{2\partial y^2}\Big|_{(x_\epsilon,y_\epsilon,t=0)}\eta_y^2+\frac{\partial^2\rho}{2\partial x^2}\Big|_{(x_\epsilon,y_\epsilon,t=0)}\eta_x^2
\\&~~~~~~~~~~~~~~~~~~~~~~~~~~~~~~~~~~~~~~~~~~~~~+\frac{\partial\rho}{\partial y}\Big|_{(x_\epsilon,y_\epsilon,t=0)}\eta_y+\frac{\partial\rho}{\partial x}\Big|_{(x_\epsilon,y_\epsilon,t=0)}\eta_x\big)d{\eta_x}d{\eta_y}
\end{split}
\end{equation}
Evaluating the Gaussian integral and using $A = \sqrt{\frac{-2\epsilon \hbar \pi i}{m}}\sqrt{\frac{2\epsilon \hbar \pi i}{m}} $ we get
\begin{equation}
\begin{split}
\rho(x_\epsilon,y_\epsilon,\epsilon)& =\rho(x_\epsilon,y_\epsilon,t=0)-\frac{i}{\hbar}V(x)\rho(x_\epsilon,y_\epsilon,t=0)\epsilon+\frac{i}{\hbar}V(y)\rho(x_\epsilon,y_\epsilon,t=0)\epsilon\\
&~~~~~~~~~~~~~~~~~~~~~~~~~~~+\frac{-i\hbar}{2m}\frac{\partial^2\rho}{\partial y^2}\Big|_{(x_\epsilon,y_\epsilon,t=0)}\epsilon+\frac{i\hbar}{2m}\frac{\partial^2\rho}{\partial x^2}\Big|_{(x_\epsilon,y_\epsilon,t=0)}\epsilon\\
& =\rho(x_\epsilon,y_\epsilon,t=0)-\frac{i}{\hbar}[H,\rho]\epsilon
\end{split}
\end{equation}which describes how a density operator evolves in time:
\begin{equation}
\frac{d{\rho}}{\dd{t}}=-\frac{i}{\hbar}[H,\rho]
\end{equation}
The above equation is the von Neumann equation and it describes the statistical state of a system in quantum mechanics. We refer to the above equation as the statistical quantum limit of GRW model.
\subsection{Classical Limit}
The following analysis is previously done by Ajanapon \cite{Pimon_1987} for the propagator of the density matrix in standard quantum mechanics. We here make use of the same analysis for the propagator of the GRW model. From equations (\ref{eq_path_int_shlok}) or (\ref{eq_path_int_bhavya}), the path integral for GRW model could be written as,
{\begin{align}
\label{eq_GRW_path_int_prop_2}
\rho(x_T,y_T,T)=&\int_{\rm all \ paths} \exp\left[\frac{i}{\hbar }\left( ~S(x_t,T,t=0) - S(y_t,T,t=0)~ \right)\right]\exp\left[- \lambda\int_0^T (1 - e^\frac{-(x_t-y_t)^2}{4r_C^2})dt \right]\nonumber\\&~~~~~~~~~~~~~~~~~~~~~~~~~~~~~~~~~~~~~~~~~\rho(x_0,y_0,t=0) [Dx_t][Dy_t] dx_0 dy_0
\end{align}}
Now we consider the limit $\lambda T \gg 1$ which could be interpreted as  waiting for a sufficiently long time, or the collapse rate $\lambda$ for the system is sufficiently large. Large $\lambda$ implies large mass since the collapse rate is directly proportional to number of entangled particles in the system. As a result large $\lambda$ implies large action. On the other hand, large time also results in large action. As a result, large masses and large times are both representatives of classical limit which causes $S$ to be large, and thus implies the limit $S \gg \hbar$.\\

When a rapidly oscillating function is multiplied with a smooth function then the integral of their product could be approximated by the smooth function at the stationary point of the rapidly oscillating function. This is commonly called the stationary phase approximation. Here $x^{cl}_t$ and $y^{cl}_t$ are the stationary paths for $S(x_t,T,t=0)$ and $S(y_t,T,t=0)$ respectively in the limit $S \gg \hbar$. Thus the stationary phase approximation leads us to the following equation,
{\begin{align}\label{eq_GRW_stat_path}
\rho(x_T,y_T,T)= & \int \exp\left[\frac{i}{\hbar }\left( ~S(x^{cl}_t,T,t=0) - S(y^{cl}_t,T,t=0)~ \right)\right]\exp\left[- \lambda\int_0^T (1 - e^\frac{-(x^{cl}_t-y^{cl}_t)^2}{4r_C^2})dt \right] 
 \nonumber \\ & ~~~~~~~~~~~~~~~~~~~~~~~~~~~~~~~~~~~~~~~~ \rho(x_0,y_0,t=0) dx_0 dy_0
\end{align}}
For brevity, we here drop the notation for stationary paths and use $x^{cl}_t= x_t$ and $y^{cl}_{t} = y_t$.
The $\rho(x_T,y_T,T)$ in the above expression represents diagonal as well as off-diagonal terms in position basis \cite{Pearle}. Now we look for the off-diagonal terms of the final $\rho$, which are specified by large $(x_t-y_t)$. In the limit $(x_t-y_t) \gg r_C$, the non-oscillating part of the propagator could be approximated as, 
	\begin{equation}
	\exp\left[- \lambda\int_0^T (1 - e^\frac{-(x_t-y_t)^2}{4r_C^2})dt \right] \approx \exp\left[- \lambda T \right]
	\end{equation}
	This leads to damping of the off-diagonal terms of the density matrix. Thus, in the limit $\lambda T \gg 1$, the integral can be considered to be vanish. This could also be interpreted as destruction of interference in the system as the off-diagonal terms are the primary representatives of interference. Now let us consider the diagonal terms of the final $\rho$, specified by $(x_t-y_t) \approx 0$. In the limit $(x_t-y_t) \ll r_C$, the non-oscillating part of the propagator could be approximated as,
	\begin{equation}
	\exp\left[- \lambda\int_0^T (1 - e^\frac{-(x_t-y_t)^2}{4r_C^2})dt \right] \approx 1
	\end{equation}
Now, we consider an infinitesimal time step $\epsilon$. 
\begin{align}
S(x_t,\epsilon,t=0) -& S(y_t,\epsilon,t=0) \nonumber\\
& =  \frac{m}{2\epsilon^2}(x_\epsilon-x_0)^2\epsilon - \frac{1}{2}\left[V(x_\epsilon)+V(x_0)\right]\epsilon - \frac{m}{2\epsilon^2}(y_\epsilon-y_0)^2\epsilon +  \frac{1}{2}\left[V(y_\epsilon)+V(y_0)\right]\epsilon \\
& =  \frac{m}{\epsilon}\left[ \frac{1}{2} (x_\epsilon + y_\epsilon) - \frac{1}{2} (x_0 + y_0) \right]~\left[(x_\epsilon - y_\epsilon)-(x_0 - y_0)\right]\nonumber \\&~~~~~~~~~~~~~~~~~~~~~~~~~~~~~~- \frac{\epsilon}{2} \left[V(x_\epsilon) - V(y_\epsilon)\right] - \frac{\epsilon}{2} \left[V(x_0) - V(y_0)\right]
\end{align}
 Motivated by the above expression, we implement the following change of variables, 
\begin{align}
\bar{q}_t &= \frac{1}{2} (x_t + y_t) \\
\Delta_t &= (x_t - y_t)\rule{0pt}{4ex}\\
U(\bar{q}_t,\Delta_t) &= V(\bar{q}_t + \frac{1}{2}\Delta_t) - V(\bar{q}_t - \frac{1}{2}\Delta_t)
\end{align}
Thus the equation (\ref{eq_GRW_stat_path}) could be written as,
	\begin{align}\label{eq_GRW_infin_time}
	\rho(\bar{q}_\epsilon,\Delta_\epsilon,\epsilon)= A\int \exp\left[\frac{i}{\hbar}\left( \frac{m}{\epsilon} (\bar{q}_\epsilon -\bar{q}_0)(\Delta_\epsilon - \Delta_0) -\frac{\epsilon}{2} U(\bar{q}_\epsilon,\Delta_\epsilon) - \frac{\epsilon}{2} U(\bar{q}_0,\Delta_0) \right)\right]\nonumber\\~~~~~~~~~~~~~~~~~~~~~~~~~~~~~~~~~~ \rho(\bar{q}_0,\Delta_0,t=0) dx_0 dy_0 
	\end{align}
As the state of a system is specified by position and momentum in classical mechanics, we take the Fourier transform of $\Delta$ as given by,
\begin{equation}
\rho(\bar{q}_t,p_t,t) = A \int e^{(-ip_t \Delta_t)} \rho(\bar{q}_t,\Delta_t,t) d\Delta_t
\end{equation}
Thus the equation (\ref{eq_GRW_infin_time}) in terms of $p_t$ could be written as,
\begin{align}\label{eq_QM_infin_time_daig_terms}
\rho(\bar{q}_\epsilon,p_\epsilon,\epsilon)=&A\int \exp\left[\frac{i}{\hbar}\left( \Delta_0 p_0 -\Delta_\epsilon p_\epsilon +\frac{m}{\epsilon} (\bar{q}_\epsilon -\bar{q}_0)(\Delta_\epsilon - \Delta_0) -\frac{\epsilon}{2} U(\bar{q}_\epsilon,\Delta_\epsilon) - \frac{\epsilon}{2} U(\bar{q}_0,\Delta_0) \right)\right]\nonumber\\&~~~~~~~~~~~~~~~~~~~~~~~~~~~~~~~~~~~~~~~~~~~~~~~~~~~~~\rho(\bar{q}_0,p_0,t=0) d\Delta_0 d\Delta_\epsilon dx_0 dy_0
\end{align}
The $\rho(\bar{q}_t,p_t,t)$ could be interpreted as the phase space representation of the diagonal terms of the density matrix in the limit $S\gg \hbar$. As the $\Delta_\epsilon \ll r_C$, $U(\bar{q}_t,\Delta_t)$ could be approximated by Taylor expanding and ignoring $\Delta_t^2$ and its higher orders
\begin{equation}
U(\bar{q}_t,\Delta_t) \approx \Delta_t \frac{\partial V}{\partial q}(\bar{q}_t)
\end{equation}
The equation (\ref{eq_QM_infin_time_daig_terms}) could be further simplified by using the above approximation,
\begin{align}
\rho(\bar{q}_\epsilon,p_\epsilon,\epsilon) &= \frac{1}{N'}\int \exp\left[\frac{i\Delta_0}{\hbar}\left( p_0 - \frac{m}{\epsilon}(\bar{q}_\epsilon - \bar{q}_0) - \frac{\epsilon}{2} \frac{\partial V}{\partial q}(\bar{q}_0) \right)\right] \nonumber\\ & ~~~~~~~~~~~~~~~\exp\left[\frac{- i\Delta_\epsilon m}{\hbar \epsilon}\left( \bar{q}_0 - \bar{q}_\epsilon + \frac{\epsilon}{m} p_\epsilon + \frac{\epsilon^2}{2m} \frac{\partial V}{\partial q}(\bar{q}_\epsilon) \right)\right] \rho(\bar{q}_0,p_0,t=0) d\Delta_0 d\Delta_\epsilon dx_0 dy_0 \\
&= \frac{1}{N''}\int \delta\left( p_0 - \frac{m}{\epsilon}(\bar{q}_\epsilon - \bar{q}_0) - \frac{\epsilon}{2} \frac{\partial V}{\partial q}(\bar{q}_0) \right) \delta\left( \bar{q}_0 - \bar{q}_\epsilon + \frac{\epsilon}{m} p_\epsilon + \frac{\epsilon^2}{2m} \frac{\partial V}{\partial q}(\bar{q}_\epsilon) \right)\nonumber \\&~~~~~~~~~~~~~~~~~~~~~~~~~~~~~~~~~~~~~~~~~~~~~~~~~~~~~~~~~~~~~~~\rho(\bar{q}_0,p_0,t=0)  dx_0 dy_0
\\ &= \frac{1}{N'''}\rho(\bar{q}_\epsilon -\frac{\epsilon}{m} p_\epsilon,p_\epsilon + \epsilon \frac{\partial V}{\partial q}(\bar{q}_\epsilon),t=0) 
\end{align}
The above equation could also be written as follows by changing the variables of $\rho$,
\begin{equation}
\rho\left(\bar{q}_0 + \frac{\epsilon}{m} p_0,p_0 - \epsilon \frac{\partial V}{\partial q}(\bar{q}_0),t=\epsilon\right) = \frac{1}{N'''} ~ \rho\left(\bar{q}_0,p_0,t=0\right)
\end{equation}
Now Taylor expanding the left hand side around the point ($q_0$, $p_0$, t=0) and equating orders of $\epsilon$, 
we get, at zeroth order,
\begin{equation}
N''' = 1
\end{equation}
at first order,
\begin{equation}
\frac{\partial \rho}{\partial t}\Big|_{(\bar{q}_0,p_0,t=0)} = - \frac{p_0}{m}~ \frac{\partial \rho}{\partial \bar{q}}\Big|_{(\bar{q}_0,p_0,t=0)} + \frac{\partial V}{\partial \bar{q}}\Big|_{(\bar{q}_0)} \frac{\partial \rho}{\partial p}\Big|_{(\bar{q}_0,p_0,t=0)}\\
\end{equation}
and dropping the subscript,
\begin{equation}\label{eq_Liouville}
\frac{\partial \rho}{\partial t} = - \{ \rho , H \}
\end{equation}
where $ H = \frac{1}{2m} p^2 + V(\bar{q})$. We refer to this equation (\ref{eq_Liouville}) as being the statistical classical limit of GRW. The above limit does not depend on a specific form of the initial density matrix, and hence is a phase space representation of a general density matrix following GRW evolution. 

\subsection{Absence of macroscopic position superpositions}
To summarise the discussion this far, we first developed a path integral formulation of the GRW model. We then showed that this gives us the correct quantum and classical limits. We shall now illustrate some important features of the classical limit through some examples.
Since we are taking the classical limit, we would consider large action and large number of nucleons (which implies large $\lambda$). Hence, the stationary phase approximation shown in Eq. (\ref{eq_GRW_stat_path}) would be valid. If we consider the case of a free particle, the stationary paths would be straight lines with $\dot{x}(t)=constant$.

Let us consider an initial condition that is formed by the superposition of two Gaussians separated by a macroscopic distance $|a_1-a_2|\gg r_C$. The resulting density matrix would be
\begin{equation}
\nonumber 		\rho(x_0,y_0,t=0)=\sum_{i,j=1}^{2}A_{ij}e^{-\frac{(x_0-a_i)^2}{r^2}}e^{-\frac{(y_0-a_j)^2}{r^2}}
\end{equation}
with $r\ll r_C$. Here, the coefficients $A_{ij}$ can be chosen such that the density matrix is a valid one (i.e. it has unit trace, it is positive semi-definite, and it is Hermitian). Putting this into Eq. (\ref{eq_GRW_stat_path}), we get

\begin{align} \label{ex1}
\rho(x_t,y_t,T)&=\int \exp\left[- \lambda\int_0^T (1 - e^\frac{-(x^{cl}_t-y^{cl}_t)^2}{4r_C^2})dt \right]  \exp\left[\frac{i}{\hbar }\left( ~S(x^{cl}_t,T,=0) - S(y^{cl}_t,T,t=0)~ \right)\right]\nonumber\\&~~~~~~~~~~\sum_{i,j=1}^{2}A_{ij}\exp\left[{-\frac{(x_0-a_i)^2}{r^2}}\right]\exp\left[{-\frac{(y_0-a_j)^2}{r^2}}\right]dx_0 dy_0
\end{align}

We can see that the terms of the initial density matrix $$A_{12}e^{-\frac{(x_0-a_1)^2}{r^2}}e^{-\frac{(y_0-a_2)^2}{r^2}}+A_{21}e^{-\frac{(x_0-a_2)^2}{r^2}}e^{-\frac{(y_0-a_1)^2}{r^2}}$$ would have $|x^{cl}_t-y^{cl}_t|\gg r_C$ for a large time. Hence, the final density matrix would have these terms damped exponentially as $$\exp\left(- \lambda\int_0^T (1 - e^\frac{-(x^{cl}_t-y^{cl}_t)^2}{4r_C^2})dt \right) \approx e^{-\lambda T}$$ Additionally, in the remaining terms where both paths start in the same Gaussian, the paths must finally also remain within a distance which is of the order $r_C$. Thus, the so-called off-diagonal terms are destroyed, while the approximately diagonal terms are preserved. 
 Note that the system transforms from a state with the superposition of two Gaussians to a statistical ensemble of the two Gaussians with probabilities $A_{11}$ and $A_{22}$ respectively. Note also that this statistical ensemble is different from a superposition as this represents classical probabilities which do not interfere. In this way, GRW destroys macroscopic superpositions.
%

\section{Discussion and Conclusion}

In our work we have derived the GRW propagator in two new ways. As mentioned in the introduction, the GRW propagator amounts to adding a damping term to the standard propagator that destroys macroscopic superpositions. We note that in this approach, the transition from GRW to classical and quantum mechanics is quite naturally obtained. In order to see the transition to standard quantum mechanics, we took the limit $\lambda T \ll 1$ of the path integral for the GRW model and were left with quantum mechanics for a density matrix i.e. the Von Neumann equation. In order to see the transition to classical mechanics, we took the limit $\lambda T \gg 1$ in the path integral for the GRW model and were left with the classical Liouville equation.

Our study suggests methods for generalising spontaneous localisation to the relativistic case, via the path integral representation of quantum field theory. What we see in Eqn. (23) is that spontaneous localisation is equivalent to modifying
the standard path integral by a regulator. In relativistic quantum field theory, we replace space-time coordinates by quantum fields over space-time, so that the action function $S(x,T)$ is replaced by a functional; $S(\phi(x,T))$. We propose to introduce a regulator, analogous to the one introduced in the present paper, and investigate how it might incorporate spontaneous localisation in quantum field theory.

\bibliography{GRW2}

\end{document}